%% file: ModeConverter9.tex
\documentclass[a4paper,5p]{elsarticle}
\usepackage{graphicx}
\usepackage{caption}
\usepackage{subcaption}
\usepackage{amsmath}
\usepackage{siunitx}
\usepackage[utf8]{luainputenc}
\usepackage{wasysym}
\usepackage{todonotes}

\usepackage{braket}
\usepackage[hidelinks]{hyperref}
\usepackage{tikz}

\newcommand{\op}[1]{\ensuremath{\hat{#1}}}

\newcommand{\expect}[1]
{
\langle#1\rangle
}

\begin{document}
\journal{Ultramicroscopy}
\title{$\pi/2$ Mode Converters and Vortex Generators for Electrons}
\author[ifp]{C.~Kramberger}
\author[ifp,ustem]{S.~Löffler}
\author[ustem]{T.~Schachinger}
\author[ceos]{P.~Hartel}
\author[ceos]{J.~Zach}
\author[ifp,ustem]{P.~Schattschneider\corref{cor1}}
\cortext[cor1]{Corresponding author}
\ead[cor1]{peter.schattschneider@tuwien.ac.at}

\address[ifp]{Institute of Solid State Physics, TU Wien, Wiedner Hauptstraße 8-10/E138, 1040 Wien, Austria}
\address[ustem]{University Service Center for Transmission Electron Microscopy, TU Wien, Wiedner Hauptstraße 8-10/E057-02, 1040 Wien, Austria}
\address[ceos]{CEOS Corrected Electron Optical Systems GmbH, Englerstraße 28, 69126 Heidelberg, Germany}

\begin{abstract}
In optics, mode conversion is an elegant way to switch between Hermite Gaussian and Laguerre Gaussian beam profiles and thereby impart orbital angular momentum onto the beam and to create vortices. In optics such vortex beams can be produced in a setup consisting of two identical cylinder lenses. In electron optics, quadrupole lenses can be used for the same purpose. Here we investigate generalized asymmetric designs of a quadrupole mode converter that may be realized within the constraints of existing electron microscopes and can steer the development of dedicated vortex generators for high brilliance electron vortex probes of atomic scale.
\end{abstract}



\maketitle

\section{Introduction}
A vortex beam can be characterized by a discontinuity in the phase that dictates a central void in the intensity profile. Further it features a quantized orbital angular momentum (OAM) in units of $\hbar$. The associated magnetic moment and chirality make electron vortex probes sensitive to magnetic excitations and even give them the ability to discriminate chiral crystals \cite{Juchtmans2015,Juchtmans2015a,Idrobo2016}. The development in the field was propelled by the close analogy to the established methods for optical vortex generation \cite{Bliokh2017,Lloyd2017,Larocque2018} as well as their application in helical spectroscopy \cite{Schattschneider2006,Schattschneider2014,Pohl2017}. While light optics has stimulated several methods of electron vortex generation \cite{Uchida2010,Verbeeck2010,Grillo2014}, electron vortices can also be formed by multipoles \cite{Clark2013} or magnetic fields \cite{Beche2014,Tavabi2018}. Each of these methods has its own merits and challenges, but none of them offers a pure singular OAM state without the need to block out unwanted portions of the intensity.

Yet one particular optical setup holds the promise to work on the entire beam in high purity, so that there would not be any need to filter out other diffraction orders or spurious unwanted angular states: the $\pi/2$ mode converter (MC) \cite{Beijersbergen1993} converts Hermite Gaussian (HG) beams to corresponding Laguerre Gaussians (LG) ones and \textit{vice versa}. The first order cases are:   
\begin{eqnarray}
HG(x,y) &\propto& 2x\cdot e^{-\frac{x^2+y^2}{w^2}},\label{eq:HG}\\
LG(r,\phi) &\propto& 2r\cdot e^{-\frac{r^2}{w^2}}\cdot e^{i\phi}.\label{eq:LG}
\end{eqnarray}
$(x,y)$ and $(r,\phi)$ are the Cartesian and polar coordinates, respectively. $w$ is the width defining parameter.  
While two-way LG to HG beam conversion was demonstrated in a proof of principle experiment for electron beams \cite{Schattschneider2012}, mode matching could not be achieved. Therefore the donut profile was only transient and could not be projected to another plane.

\begin{figure}[htp]
	\centering
	\includegraphics[width=\linewidth]{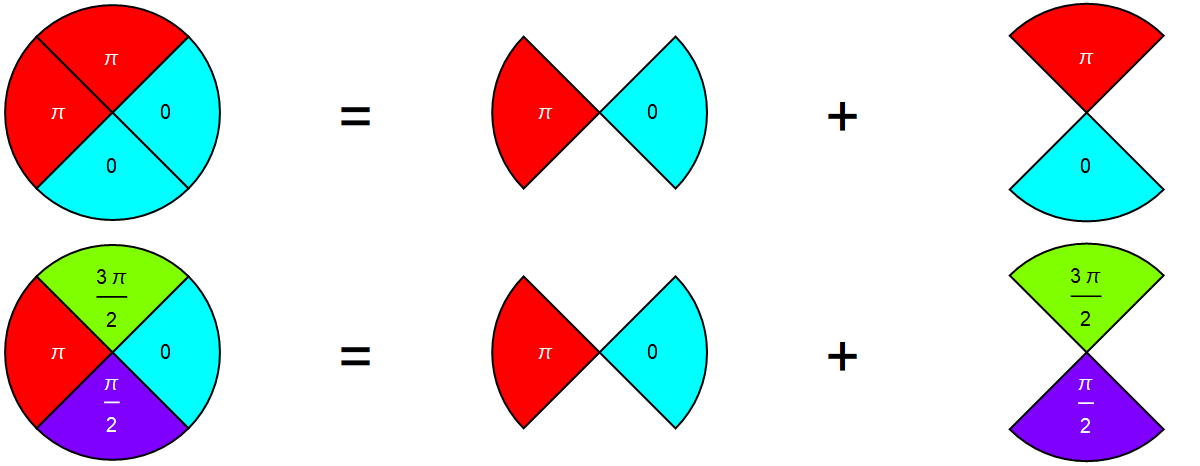}
	\caption{principle of $\pi/2$ mode conversion. The output beam of a Hilbert device can be split into a sum of two HG-like components. Acquiring a Guoy shift of $\pi/2$ in one of the components produces a stepwise azimuthal phase ramp.}
	\label{fig:splitbeam}
\end{figure}

To picture how mode conversion occurs we can replace an incident beam with a straight central phase jump of $\pi$ (a Hilbert beam) by two sub-waves that possess the same mirror symmetry as vertical and horizontal HG modes. Figure~\ref{fig:splitbeam}
demonstrates quite generally the essence of mode conversion: The two subwaves propagate independently from the entrance to the exit, where they will have accumulated a relative phase (i.e. Guoy shift) of $\pi/2$. Thus their coherent superposition creates an azimuthal \lq stair case\rq~ phase ramp and a corresponding ring current. The beam has now non-zero OAM.

We propose that the quadrupole lenses in existing aberration correctors can be re-purposed to realize a fully functional $\pi/2$ MC. When an incoming beam is prepared with a suitable wavefront pattern, it would be completely transformed into a vortex beam without sacrificing intensity. 
Switching between left and right handed helical operation would be as stable and reproducible as setting electron lenses. 
The ability to perform mode conversion on electron beams will doubtlessly also open new avenues in mode sorting \cite{Grillo2017}, especially in conjunction with programmable phase masks \cite{Verbeeck2018}. 

We present an analytical treatment of general asymmetric setups and also run simulations for entire electron optical setups of different $\pi/2$ MCs.

\section{Gaussian Mode Converters}
\subsection{The optical $\pi/2$ mode converter}
We reproduce shortly the principle of the symmetric $\pi/2$ MC given in Ref.~\cite{Beijersbergen1993}. The schematic of the setup is illustrated in Fig.~\ref{fig:OpticalModeConverter}.   There is an astigmatic beam waist located at $z=0$ and two cylinder lenses at positions $z=-a$ and $z=a$ with a distance of $d=2a$. In light optics with static lenses this distance is the only adjustable degree of freedom in the setup. Apart from the incident Gaussian beam, there are 3 conditions to be fulfilled. The conditions are:

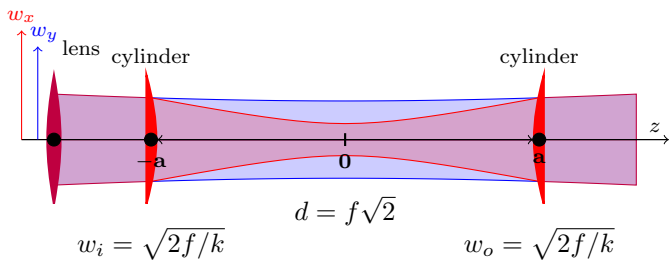
\begin{figure}[htbp]
	\centering
	\input{Figure2.tex}
	\caption{Symmetric optical $\pi/2$ mode converter. The blue shaded $yz$ component is a quasi-collimated beam. The identical cylinder lenses only act on the red shaded $xz$ component.}
	\label{fig:OpticalModeConverter}
\end{figure}

\begin{itemize}
\item{} the widths of Gaussian profile evolve as
\begin{equation}
w(z)=\sqrt{1+\left(\frac{z}{z_r}\right)^2}\cdot\sqrt{\frac{2 z_r}{k}}. 
\end{equation}
Here $z_r$ is the Rayleigh range. The wavenumber $k$ and wavelength $\lambda$ follow $k\lambda=2\pi$. The widths in the $xz$ and $yz$ cuts must be equal at the exit plane $z=+a$:
\begin{equation}
w_x(a)=w_y(a).
\label{eq:1}
\end{equation}

 \item{} the Gouy phase difference of the $xz$ and $yz$ components accumulated at the second cylinder lens must be $\pi/2$. Due to center symmetry this is equivalent to:
\begin{equation}
	\tan^{-1}\left(\frac{a}{z_{rx}}\right) -\tan^{-1}\left(\frac{a}{z_{ry}}\right)= \frac{\pi}{4}.
	\label{eq:2}
\end{equation}

\item{} the radii of curvature 
\begin{equation}
R(z)=z \left(1+\left(\frac{z_r}{z}\right)^2\right)
\end{equation}
in the $xz$ and in the $yz$ cuts after the second cylinder lens must fulfill 
\begin{equation}
R_{x,z}(a)=R_{y,z}(a)=-R(-a).
\label{eq:3}
\end{equation}
\end{itemize}
Equations \ref{eq:1} and \ref{eq:2} give immediately
\begin{equation}
z_{rx}=a\left(\sqrt{2}-1\right), \quad z_{ry}=a\left(\sqrt{2}+1\right).
\label{eq:z}
\end{equation}
With this result, the widths at the entrance and exit planes follow as:
\begin{equation}
w(-a)=w(a)=\sqrt{\frac{\sqrt{2}\cdot d}{k}}.
\label{eq:w}
\end{equation}
With Newton's equations,
\begin{equation}
\frac{1}{R_i} = \frac{1}{R_x(-a)} + \frac{1}{f} = \frac{1}{R_y(-a)},
\label{eq:newtonCylinder}
\end{equation}
and the previous results for the width Eqn.~\ref{eq:w} and the Rayleigh ranges Eqn.~\ref{eq:z},
this gives 
\begin{equation}
d=\sqrt{2}\cdot f
\label{eq:foptical}
\end{equation}
for a symmetric optical $\pi/2$ MC.

\subsection{The quadrupole $\pi/2$ mode converter}
The key differences between electron and light optics are that the distances and positions are fixed but the focal lengths can be controlled via lens excitations. Also, quadrupoles (QPs) are widely available. For electrons we thus propose to replace the cylinder lenses with QPs. This has no effect on Eqns.~\ref{eq:1}\&\ref{eq:2} and hence on the Rayleigh ranges $z_R$. If the QPs are always focusing the $xz$ component and defocusing the $yz$ component,
then Newton's equations read as:
\begin{equation}
\frac{1}{R_i} = \frac{1}{R_x(-a)} + \frac{1}{f} = \frac{1}{R_y(-a)} - \frac{1}{f}.
\label{eq:newtonQP}
\end{equation}
This modifies the relation from Eqn.~\ref{eq:foptical} to
\begin{equation}
f=\sqrt{2}\cdot d
\end{equation}
and the widths at the entrance and exit plane scale accordingly:
\begin{equation}
w(-a)=w(a)=\sqrt{\frac{2\cdot\sqrt{2}\cdot d}{ k}}.
\label{eq:waist}
\end{equation}
Notably, the curvatures of the incident and outgoing beam simplifies to
\begin{equation}
R_i= -R_o = -d.
\label{eq:Rasy}
\end{equation}
Replacing cylinder lenses with QPs simplifies the solutions of the mode matching conditions considerably. The incoming and outgoing curvatures and the quadrupole focusing (or defocusing) are functions of the distance $d$ only, and not of the wavenumber $k$. Only the beam widths scale with $\sqrt{d/k}$. The actions of round lenses and QPs correspond to two orthogonal Zernike polynomials ($Z_2^0$ and $Z_2^2$, respectively), while a cylinder lens is a superposition of the two. QPs are therefore better suited for aligning a $\pi/2$ MC.  

\subsection{The asymmetric $\pi/2$ mode converter}

If the constraint of equal focal lengths of the two QPs is relaxed, then the beam waists will be at different positions for the $xz$ and $yz$ component. In addition, the incoming and outgoing widths will differ, but the radii of curvature in Eqn.~\ref{eq:Rasy} are not affected. 

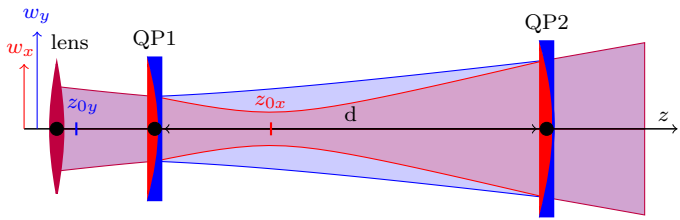
\begin{figure}[htbp]
	\centering
	\input{Figure3.tex}
	\caption{Schematics of an asymmetric $\pi/2$ MC for electrons with $w_o/w_i=2.12$ as in the simulations in Figs.~\ref{fig:phaselines}\&\ref{fig:PsiAsymSpherical}. The preceding round  lens provides the correct curvature. The quadrupoles (QP1 \& QP2) focus the $xz$ component (red shading) and defocus the $yz$ component (blue shading).}
	\label{fig:AsymModeConverter}
\end{figure}
If Eqns.~\ref{eq:1}\&\ref{eq:3} are met, (i.e. the astigmatism is canceled) then the
the relative Guoy shift $\Psi$ is given by the distance $d$ as well as the QPs focal lengths $f_i$ and $f_o$.
\begin{equation}
\tan\Psi = \frac{2u}{1-u^2},\quad u^2 = \frac{f_if_o}{d^2}-1.
\label{eq:guoy}
\end{equation} 
In a properly aligned $\pi/2$ MC, $\Psi$ is $\pi/2$ and the dimensionless parameter $u$ becomes 1. The conditions read (see supplementary information):
\begin{eqnarray}
 f_i\cdot f_o = 2\cdot d^2, \label{eq:fasy}\\
 w_i=\sqrt{\frac{2\cdot f_i}{k}}, \quad\quad\quad w_o=\sqrt{\frac{2\cdot f_o}{k}}. \label{eq:wasy}
\end{eqnarray}
The proper choice of quadrupole focal lengths Eqn.~\ref{eq:fasy} for an incoming beam width $w_i$  according to Eqn.~\ref{eq:wasy} allows to achieve $u = 1$ with a magnification of $w_o/w_i$. A schematic example is sketched in Fig.~\ref{fig:AsymModeConverter}.

\subsection{Practical considerations}

When it comes to electron optical alignment, the very appealing benefit of the asymmetric $\pi/2$ MC design is, that there is only one prior requirement on the non-astigmatic incoming beam. Its curvature has to be centered onto the principal plane of QP2. This can be readily achieved by focusing a wide enough beam onto QP2. If the incoming  width $w_i$ is several times larger than the width $w$ that would be required in a symmetric $\pi/2$ MC (Eqn.~\ref{eq:waist}), $z_r << d$ will also hold, and the required lens excitations can be found by minimizing the effects of wobbling QP2. Then a smaller condenser aperture can be used with the same lens settings, to provide a smaller $w_i$ with the correct curvature. If the reduced $w_i$ is comparable to the $w$ of the symmetric $\pi/2$ MC, the required $f_i$ and $f_o$ will also be comparable. Since the outgoing radius of curvature does not depend on the Guoyshift, pairs of $f_i$ and $f_o$ can be realized by choosing any one and adjusting the other one, until $u$ from Eqn.~\ref{eq:guoy} becomes 1. There is no need to match the width of the symmetric $\pi/2$ MC exactly.

\section{Spherical Mode Converters}
\subsection{Guoy phase}

\begin{figure}[hbt]
	\centering
	\includegraphics[width=\linewidth]{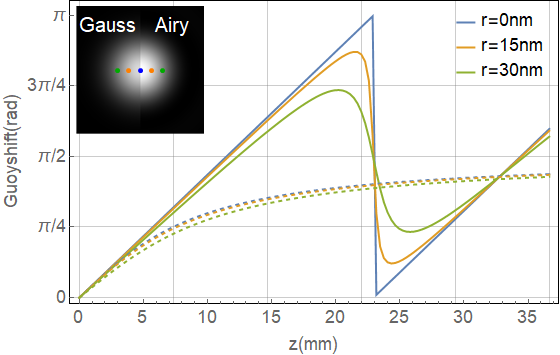}
	\caption{Guoy phase of a spherical wave (full lines) and an HG$_{00}$ beam (dashed) at different radii.  A lens with a focal length of 120 mm is positioned at $z=-119.4$~mm. The aperture radius and Gaussian width at the lens are $w=r/\sqrt{2}=1250$~nm. Acceleration voltage $U_a=200$~kV. The inset shows the Gaussian waist and Airy disk formed at $z=0$, the dots correspond to the different radii. Horizontal gridlines are at $\pi/4$, $\pi/2$ and $3\pi/4$, vertical gridlines count Rayleigh ranges $z_r=7.3$~mm of the Gaussian beam.}
	\label{fig:GuoyPhase}
\end{figure}

Sculpting a Gaussian intensity profile is impractical if not impossible in  electron microscopy. Instead we consider standard spherical waves as an input. The Guoy phase of an astigmatic higher order HG$_{nm}$ beam follows a $\tan^{-1}$ function.
\begin{equation}
\Psi=(n+\frac{1}{2}) \tan^{-1}(\frac{z-z_x}{z_{rx}})
+(m+\frac{1}{2}) \tan^{-1}(\frac{z-z_y}{z_{ry}})
\label{eq:guoynm}
\end{equation}
where $z_{x}$ and $z_{y}$ are the positions of the line foci and $z_{rx}$ and $z_{ry}$ are the respective Rayleigh lengths. A spherical wave has a different Guoy phase. A typical example of an incoming electron beam in the geometric optic regime is shown in Fig.~\ref{fig:GuoyPhase}.

It is linear around zero defocus \cite{Born1999}, and for small defocus values up to one Rayleigh range it approximates well that of a HG$_{00}$ beam with the same focus under the condition that the aperture radius is $\sqrt{2}$ times as large as the width $w$ of the incoming Gaussian.  The traces for the Guoy phase at roughly 1/3 and 2/3 of the radius of the Airy disk are also very linear up to $z=z_R$. This comparison illustrates that a Gaussian input for the $\pi/2$ MC can be replaced by a spherical wave with a scaled diameter.

\subsection{Hilbert beams}

A feasible approach to produce an electron beam similar to a HG is a Hilbert plate that induces a phase shift of $\pi$ between the two halfs of a round aperture. One may also use a magnetic bar to this aim \cite{Tanji2015,Guzzinati2017}. In the following, we shall refer to such a phase shifter as a Hilbert device, independent of the principle used. Beams produced with such a device are henceforth called spherical Hilbert beams.

The Guoy shift of HG beams in the mode converter can be calculated analytically with Eqn.~\ref{eq:guoynm}. For spherical Hilbert beams we have to resort to wave optical simulations. 
To this aim we performed two independent simulations of an asymmetric QP $\pi/2$ setup for the horizontal and vertical components as suggested in Fig.~\ref{fig:splitbeam}. 
The magnification is the same as in Figs.~\ref{fig:phaselines}\&\ref{fig:PsiAsymSpherical} with $r_i = 357$~nm, $d=120$~mm and quadrupole focal lengths of $f_i = 80$~mm and $f_o= 360$~mm. The acceleration voltage is $U_a=200$~kV. 

\begin{figure}[htb]
	\centering
		\includegraphics[width=\linewidth]{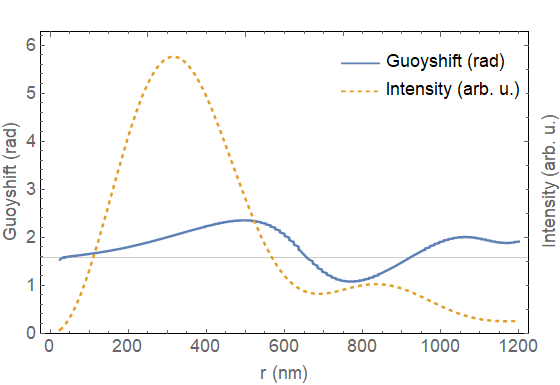}
	\caption{Solid: numerical Gouy shift for a split Hilbert beam (see Fig.~\ref{fig:splitbeam}) after a $\pi/2$ MC. The horizontal grid line marks the targeted phase shift of $\pi/2$. The shared dashed radial intensity profile is taken along the mirror axis.}
	\label{fig:GuoyShift}
\end{figure}

The propagated components were rotated so that their symmetry axis coincide. The difference in Guoy phase is traced in the direction of the aligned symmetry axis. The Guoy shift obtained in that way oscillates around the ideal value of $\pi/2$ that would be the outcome for an ideal HG input. The radial intensity profile shows that an acceptable average Guoy shift can be achieved at the most relevant radius.

\subsection{Orbital angular momentum}
\begin{figure}[htb]
	\centering
		\includegraphics[width=0.48\linewidth]{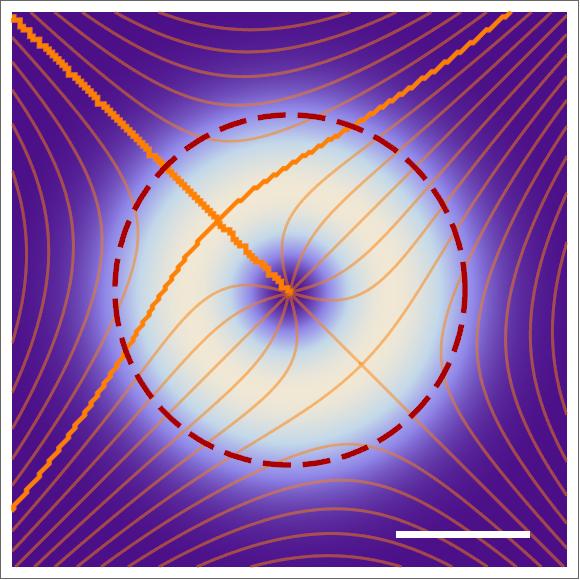}
		\includegraphics[width=0.48\linewidth]{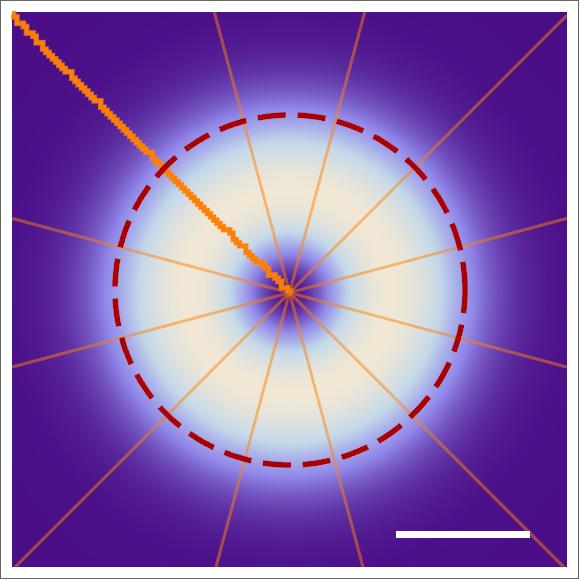}\\
		\includegraphics[width=0.48\linewidth]{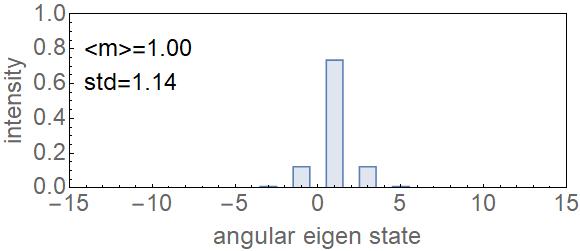}
		\includegraphics[width=0.48\linewidth]{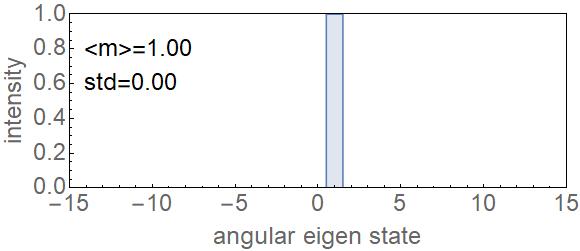}\\	
		\includegraphics[width=0.48\linewidth]{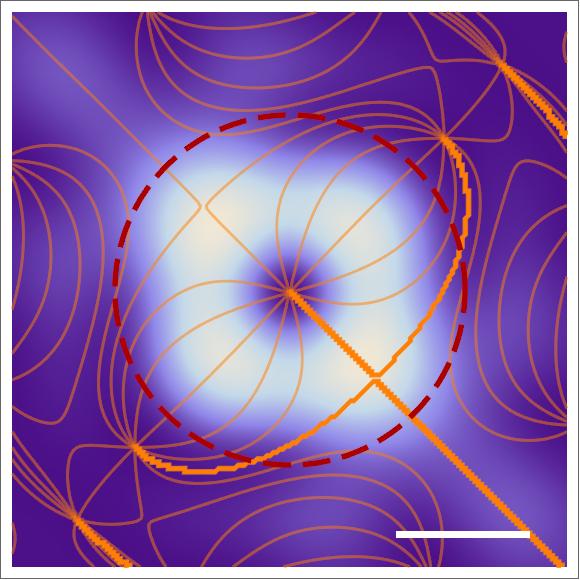}
		\includegraphics[width=0.48\linewidth]{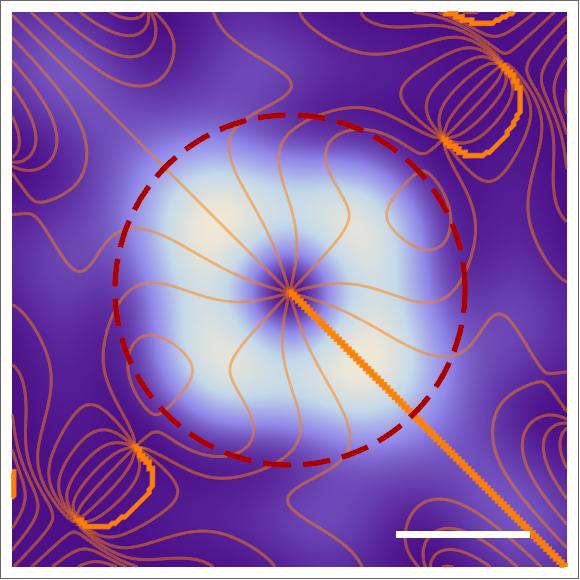}\\
		\includegraphics[width=0.48\linewidth]{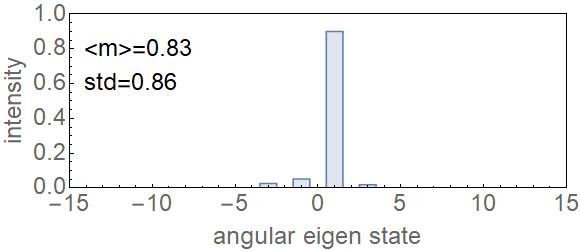}
		\includegraphics[width=0.48\linewidth]{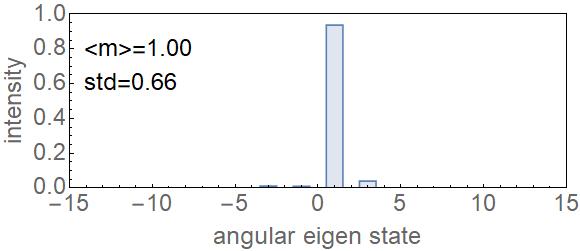}
	\caption{Vortex beams produced by a $\pi/2$ MC.
		Isophasal lines are superimposed on the intensities and angular spectra immediately before (left) and after (right) QP2 in the same setup as in Fig.~\ref{fig:GuoyShift}.  Upper row: HG beam with width $w_o=537$~nm, Lower row: spherical Hilbert beam with radius $r_o=759$~nm.  Scale bar: 500 nm. The dashed circle ($r=650$~nm) marks the area selected for decomposition into azimuthal eigenmodes.}
	\label{fig:phaselines}
\end{figure}

When expanding the wave function of the propagating beam $\psi$ in a given plane into $\op L_z$ eigenfunctions
\begin{equation}
\psi(r, \phi)= \sum_m c_m(r) e^{i m \phi},
\label{eq:cl0}
\end{equation}
the expectation value of the OAM can be calculated as
\begin{equation}
\expect{\op L_z}=\frac{\bra{\psi}  \op L_z  \ket{\psi}}{\braket{\psi |\psi}}
= \hbar \, \frac{\sum_m m \int{|c_m(r)|^2 \, r \, dr}}{\sum_m  \int{|c_m(r)|^2 \,r \, dr}}.
\label{eq:cl}
\end{equation}
Figure~\ref{fig:phaselines} shows the phase structure before and after the second quadrupole for a $HG_{0,1}$ and a Hilbert beam. The parameters for the $\pi/2$ MC are identical to those in Figs.~\ref{fig:GuoyShift}\&\ref{fig:PsiAsymSpherical}.
Note that the phase structure has been compensated for the diverging curvature Eqn.~\ref{eq:Rasy}. The remaining purely azimuthal phase structure at the entrance to QP2 is visibly astigmatic for both beam profiles. Indeed, the decomposition according to Eqn.~\ref{eq:cl0} reveals a broadened distribution. After QP2 the astigmatism is corrected, and the $m=1$ contribution increases. The $HG_{0,1}$ beam is transformed into a clean $m=1$ LG state. QP1 did already exert the full torque of $\expect{m}=1$, while QP2 establishes mode purity. The Hilbert beam picks up angular momentum on QP1 and QP2 and does also acquire a total of $\expect{m}=1$, albeit with a slightly lower $m=1$ mode purity. The mode purity may be further increased by another aperture, as the central region shows an ideal linear azimuthal phase spiral. 

\section{Numerical Simulations}

So far the setups for $\pi/2$ MCs were very much simplified. They were modeled by a composite input of an aperture, a Hilbert device, a lens and a QP followed by one single propagation step and a composite output of a lens and a QP. The analytic treatment of Gaussian beams passing through such stylized $\pi/2$ MCs as well as the very similar behavior of HG and Hilbert beams in test scenarios suggest that vortex generation is possible in an actual aberration corrected TEM. Numerical simulations on more realistic and complete setups are indispensable to confirm and possibly retune the parameters for real world electron optical designs. The obvious challenge of simulating an entire electron optical setup, is keeping track of multiple optical devices and propagation steps in between them.     

\subsection{Rescaled propagation}
In an extended optical system, like an entire TEM column, different sections of the beam have very different lateral extend or magnification. One very efficient way to adapt the lateral scale to a propagating beam can be to replace the combined action of a lens with focal length $f$ and further propagation over a distance $d$ with the combined action of a propagation over a distance $d'$, a lateral rescaling and a lens with focal length $f'$.
So instead of propagating forward to the imaging plane, the incident wavefront is propagated backwards to the object plane. Then the magnification of the imaging and a new lens with focal length $f'$ are applied. The transformed distance $d$ does not need to be the full distance to the next lens or aperture. In fact it can be chosen freely, and the signs of the rescaled and remaining distance are arbitrary. With the introduction of $s$ and $s'$ for the original and the re-scaled grid resolution, the transformations can be written as:   
\begin{eqnarray}
\frac{1}{d'} = \frac{1}{d} - \frac{1}{f}\\
\frac{s'}{s} = 1-\frac{d}{f}\\ 
f' = f-d.
\label{eq:rescaling}
\end{eqnarray}
Sign changes in $d$ are equivalent to propagating backwards. Sign changes in $f$ and $s$ trigger a mirror inversion and a phase shift of $\pi$.

\subsection{Virtual microscope}
In this section, we present a detailed numerical study of the propagation behavior of HG and Hilbert beams through an ensemble of lenses and QPs. To this aim, we have developed a {\sc JAVA} plugin for ImageJ. The graphical user interface represents a fully editable virtual microscope. Different setups can be stored in human readable and editable xml files which define among other parameters a unique order in which lenses, apertures and propagation distances are applied to an initial plane wave. Wavefronts of the propagated beam can be viewed as stacks of images.

Numerically, lenses $\mathcal{L}$, apertures, quadrupoles $\mathcal{QP}$ and other devices are represented as a complex map for the real and the imaginary part of their action, respectively.   
\begin{eqnarray}
\mathcal{L}(f,x,y) =\exp\left[i\cdot \left(x^2+y^2\right)\cdot\frac{k}{2\cdot f} \right]\\
\mathcal{QP}(f,x,y)=\exp\left[i\cdot \left(x^2-y^2\right)\cdot\frac{k}{2\cdot f}\right]
\end{eqnarray}
The simulation starts with a plane wave with phase 0 at $z=0$. At every plane, the current cross section $\psi_z$ is multiplied with the complex sheets at this plane. This step can account for arbitrary apertures, gratings,  Hilbert devices, wavefront deformations by lenses and multipoles. It can also define for instance a Hermite Gaussian. Then the wavefront is propagated through free space to the next plane.
The propagation over a distance $d$ from $\psi_z$ to $\psi_{z+d}$ can be individually configured to be carried out in customizable steps. We always employ the par-axial approximation, since lateral dimensions are $\mu$m and relevant distances are at least mm. Each step can be propagated in frequency or spatial domain.   
\begin{eqnarray}
\psi_{z+d}=\mathcal{FT}^{-1}\left(\mathcal{P}(d)\cdot\mathcal{FT}(\psi_z)\right)\label{eq:fresnel}\\ 
\psi_{z+d}(x,y) = -i\cdot\int_{x',y'}\psi_z(x',y')\cdot\label{eq:rsp}\nonumber\\
\cdot\exp\left[i\frac{(x-x')^2+(y-y')^2}{2\cdot d}\right]\,dx'dy'
\label{eq:propagating}
\end{eqnarray}
$\mathcal{FT}$ denotes Fourier transformation and the propagator $\mathcal{P}$ in Eqn.~\ref{eq:fresnel} is defined in frequency range $\hat{x},\hat{y}$  
\begin{equation}
\mathcal{P}(d,\hat{x},\hat{y}) = \exp\left[ i\cdot(\hat{x}^2+\hat{y}^2)\cdot \frac{d}{2\cdot k}\right]\\
\end{equation}
Propagation steps in spatial domain (Eqn.~\ref{eq:rsp}) may also contain a custom zoom between the planes at $z$ and $z+d$.  
  
The custom splitting and scaling and per step choice between spatial and frequency domain are found to be versatile in circumventing the need for excessive oversizing or oversampling of the complex sheets and wavefronts. All simulations could be carried out on a grid of 512x512 pixels with dynamic resolution. The phase information is consistent with the Guoy shift, but there is an arbitrary global phase factor for different planes. 
\begin{figure}[htbp]
	\centering
	\includegraphics[width=0.9\linewidth, decodearray=1 0 1 0 1 0]{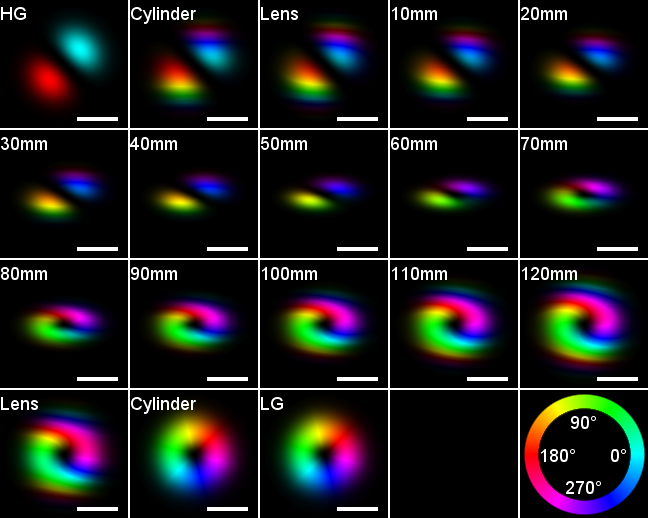}
	\caption{Propagation of an incident rotated $HG_{0,1}$ with width $w=367$~nm and  $U_a =200$~kV through cylinder lenses with $f=84.9$~mm and round lenses with $f = 409.7$~mm. The last frame is a Laguerre Gaussian (LG) with the same $w$. The scalebar is 500~nm. The wavefront is shown for every 10 mm.}
	\label{fig:PsiSymGaussian}
\end{figure}

The first test case for the virtual microscope is the symmetric cylinder lens setup.  Figure~\ref{fig:PsiSymGaussian} illustrates the propagation of the phase colored wavefronts through a basic symmetric $\pi/2$ MC setup. The incoming rotated HG$_{1,0}$ ($U_a=200$kV) has a width $w_i=367$~nm. The round lenses with $f_L = 409.7$~mm are on the inside but share the same plane with the cylinder lenses with $f_c = 84.9$~mm which are $120$~mm apart. In this ordering the effects of the first and second cylinder lens are not obscured by the isotropic curvature. The first cylinder lens introduces horizontal bands and a vertical phase curvature. The following intermediate wave fronts visualize the continuous mode conversion. And finally the second cylinder transforms the asymmetric phase pattern after the second lens into the exact LG pattern for $m=1$.

\subsection{Spherical waves and multi-scale simulation}
Moving towards a more realistic virtual setup necessitates to include round apertures and spherical waves as well as the condenser and the objective lens systems. 

\begin{figure}[htb]
	\centering
	\input{Figure8.tex}
	\caption{Complete optical setup for vortex generation. This setup is used in Fig.~\ref{fig:PsiAsymSpherical}. Lenses (purple) and quadrupoles (red/blue) focus an incoming Hilbert beam. Black and gray dots mark real images, the red/blue dot mark the real/virtual astigmatic line focus of the first quadrupole. The given diameters $\diameter$ are according to geometric optics.}
	\label{fig:entireSetup}
\end{figure}
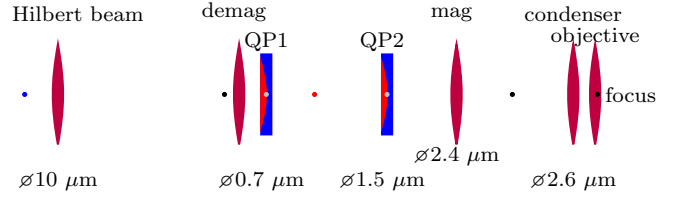

The full asymmetric $\pi/2$ MC setup is sketched in Fig.~\ref{fig:entireSetup}. The Hilbert device is assumed to be mounted in the condenser system and has a diameter of $10~\mu$m. The black dots mark actual images. The lens labeled "demag" would form another image (gray dot) at the principal plane of the second QP (QP2). The first QP (QP1) introduces astigmatism and forms one real line focus (red dot). The corresponding perpendicular line focus is virtual (blue dot). After the second QP (QP2) the beam is mode matched and appears as if emanating from an image (gray dot) in the principal plane of QP1. It is refocused in another real image in front of the condenser/objective system. The last black dot is in the focus of the objective.

\begin{figure}[htb]
	\centering
	\includegraphics[width=\linewidth, decodearray=1 0 1 0 1 0]{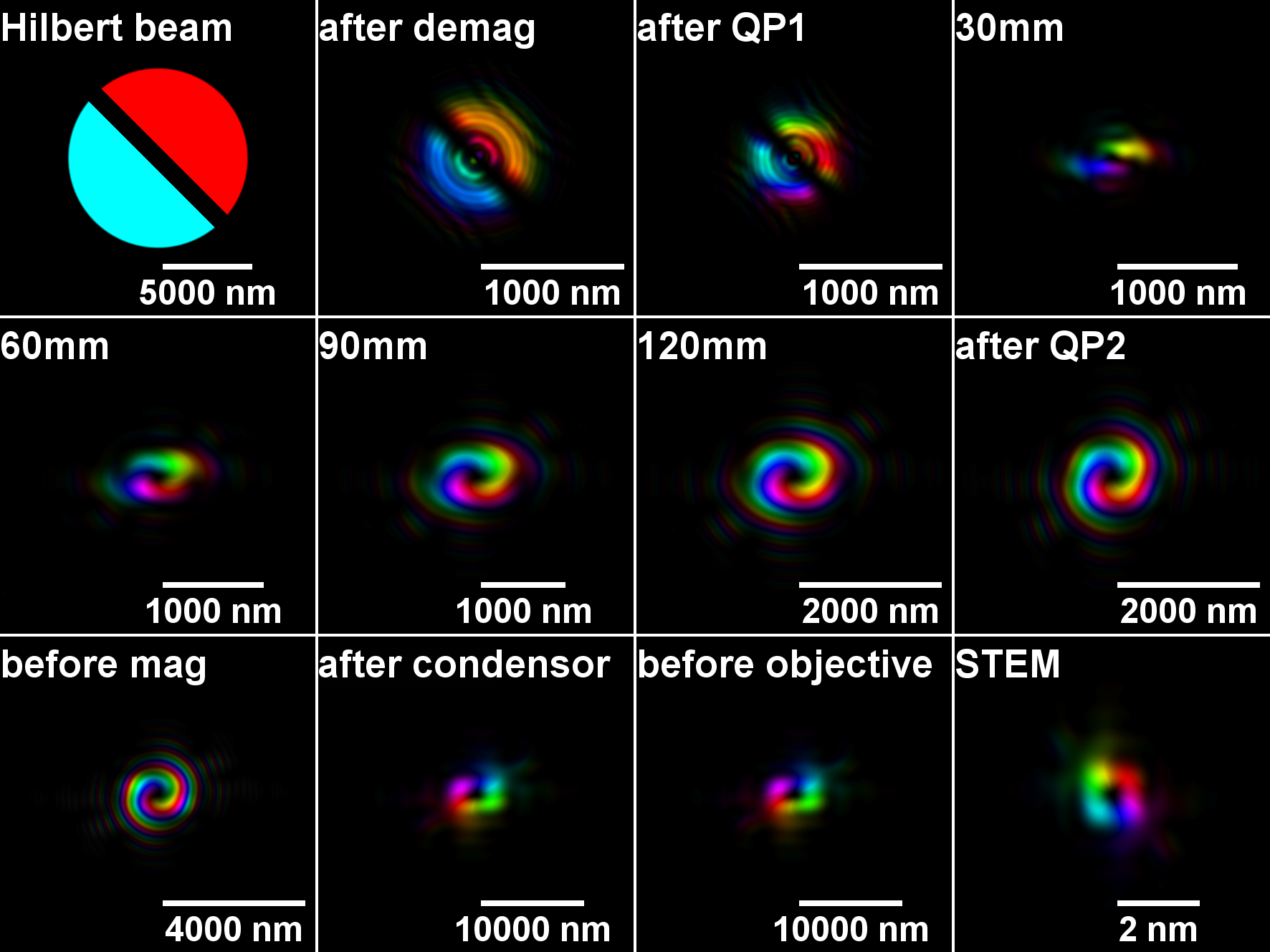}
	\caption{Propagation from a Hilbert device in the condenser to a STEM probe with orbital angular momentum $\pm\hbar$. Captions are explained in the text. The hue coloring is identical to Fig.~\ref{fig:PsiSymGaussian}}
	\label{fig:PsiAsymSpherical}
\end{figure}

The full wave optical simulation for the extended asymmetric $\pi/2$ MC setup in Fig.~\ref{fig:entireSetup} is shown in Fig.~\ref{fig:PsiAsymSpherical}. The simulation takes advantage of variable resolution on a fixed size grid. It starts from a Hilbert device in an aperture with a diameter of 10~$\mathrm{\mu}$m. A 1~$\mathrm{\mu}$m wide magnetic bar induces a phaseshift of $\pi$ between the two sides. Similar devices have been demonstrated \cite{Tanji2015,Guzzinati2017}. At the exit of the lens "demag" the wavefront is demagnified and rather reminiscent of a HG beam. The demagnification at QP1 is $14.0$ fold and the center of the converging curvature is at QP2 (Eqn.~\ref{eq:Rasy}). The distance $d$ between the QPs is $120$~mm. The quadrupoles are excited asymmetrically with $f_i=80$~mm and $f_o=360$~mm to match the incoming width (Eqn.~\ref{eq:wasy}) and to provide the correct Guoy shift (Eqn.~\ref{eq:guoy}). The output of the $\pi/2$ MC is clearly a vortex beam and the spiraling phase pattern has a diverging curvature centered at QP1. The magnification inside the $\pi/2$ MC is $w_o/w_i = 2.12$. The next frames are before the "mag" lens and after the condenser. The magnified beam is focused by the objective lens to form a donut shaped STEM probe. Changing the helicity of the STEM probe is as straight forward and reproducible as rotating the QPs by $90^\circ$, which is an crucial aspect for measuring dichroism \cite{Schattschneider2012a}.

\begin{figure}[htb]
	\centering
	\begin{minipage}[b]{0.48\linewidth}
	\includegraphics[width=\linewidth]{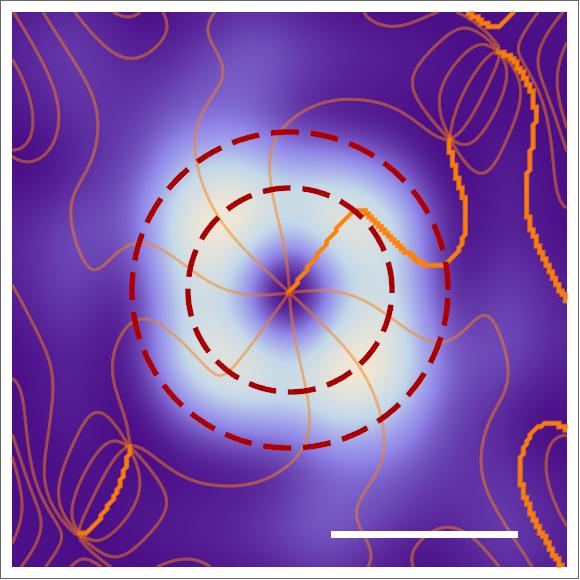}
	\end{minipage}
	\hfill
	\begin{minipage}[b]{0.48\linewidth}
	\includegraphics[width=\linewidth]{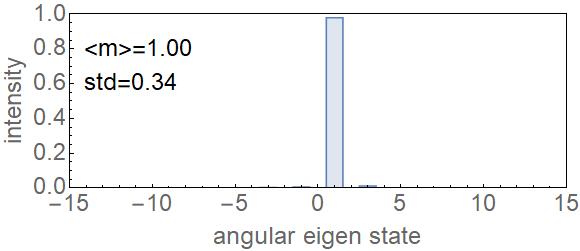}
	\includegraphics[width=\linewidth]{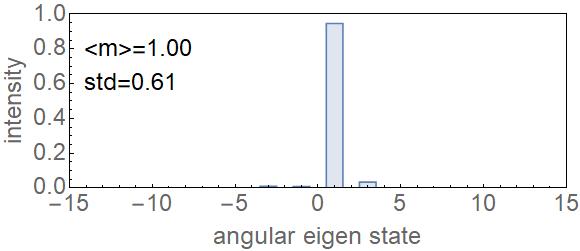}
	\end{minipage}
	\caption{Isophasal contours and angular mode distribution of the same STEM probe as in Fig.~\ref{fig:PsiAsymSpherical}. The scalebar is 1~nm. The upper spectrum is for the beam inside the smaller circle ($r=0.55$~nm) and the lower spectrum is for the beam inside the bigger circle ($r=0.85$~nm)}
	\label{fig:SphericalFocus}
\end{figure}

Notably the central bar in the Hilbert device and its diffractive blurring upon propagation to the first QP contribute to the resemblance of a HG$_{1,0}$ beam. Except for the magnification the cross sections from the interior of the $\pi/2$ MC closely resemble the internal cross sections shown in Fig.~\ref{fig:PsiSymGaussian}. 
The differences in the spiraling phase pattern before (120 mm) and after QP2 might seem subtle in direct phase coloring, but they are the same as in the isophasal representation shown in Fig.~\ref{fig:phaselines}. The second QP is crucial for canceling the astigmatism introduced by QP1 and hence stabilizing the vortex state. The donut profile is then magnified and focused by the objective lens. This is very well demonstrated by the collection of vortex profiles at various magnifications in Fig.~\ref{fig:PsiAsymSpherical}.

The resulting STEM probe is shown in  more detail in Fig.~\ref{fig:SphericalFocus}. Its azimuthal spectra show a very similar mode distribution for the two selected different radii. More importantly, both spectra clearly show that the vortex beam is stable and can be projected all the way from the second QP to the focus of the objective lens. 

We assumed realistic distances and respected minimum focal lengths for the simulation in Fig.~\ref{fig:PsiAsymSpherical}. So the predicted donut diameter of $\sim1$~nm at $\sim$2~mrad should be achievable in existing electron microscopes. We confirmed numerically that the effects of spherical aberrations $C_s$ with a typical value of a few mm and a finite source size of 50~nm are not detrimental to the predicted donut probe.  

\section{Conclusion}
We have explored the realm of possible designs for $\pi/2$ MCs in electron optics based on reconfiguring well established and relatively wide spread probe correctors. Using already existing quadrupoles is an appealing aspect, because there is no need for mechanical modifications of the TEM column and helicity switching would be straight forward. The most relevant parameters are the distance between the two quadrupoles $d$, the possible excitations of the quadrupoles or minimal $f_i$ and $f_o$, as well as the incident virtual aperture size. Allowing for asymmetry in the quadrupole excitations introduces a magnification or de-magnification and leads to an effective decoupling of the constraints on achieving isotropic width and curvature as well as $\pi/2$ mode conversion at the exit plane. 
We propose that a $\pi/2$ MC can be used to generate a very pure, and switchable $m=\pm1$ vortex beam. The design of the Hilbert device we have considered here numerically is minimalistic, and there are conceivable aperture designs that could mimic a HG$_{0,1}$ input beam even more closely.   
Significantly smaller probe diameters could be envisaged in dedicated setups with intermediate magnification stages and additional apertures.

\section*{Acknowledgements}
CK \& PS acknowledge financial support of the Austrian Science Fund (FWF): P29687-N36. TS acknowledges financial support of the Austrian Academy of Sciences: DOC-scholarship.


\end{document}

%% file: Figure2.tex
\begin{tikzpicture}[x=0.00096\linewidth,y=0.00096\linewidth]
\footnotesize
\fill[blue, fill opacity=0.2, draw=blue, domain=200:800, variable=\x] 
(200,200)
--plot({\x}, {200 + 60.1774*sqrt(1+pow((\x-500)/724.264,2))})
--(800,200)
--(200,200)
--plot({\x}, {200 - 60.1774*sqrt(1+pow((\x-500)/724.264,2))})
--(800,200)
--cycle;

\fill[red, fill opacity=0.2, draw=red, domain=200:800, variable=\x] 
(200,200)
--plot({\x}, {200 + 24.9263*sqrt(1+pow((\x-500)/124.264,2))})
--(800,200)
--(200,200)
--plot({\x}, {200 - 24.9263*sqrt(1+pow((\x-500)/124.264,2))})
--(800,200)
--cycle;

\fill[red!50!blue, fill opacity=0.36, draw=purple, domain=50:200] 
(50,200)
--plot (\x, {200 + 60.1774*sqrt(1+pow((\x-500)/724.264,2))})
--(200,200)
--(50,200) 
--plot (\x, {200 - 60.1774*sqrt(1+pow((\x-500)/724.264,2))})
--(200,200)
--cycle;

\fill[red!50!blue, fill opacity=0.36, draw=purple, domain=800:950]
(800,200)
--plot (\x, {200 + 60.1774*sqrt(1+pow((\x-500)/724.264,2))})
--(950,200)
--(800,200)
--plot (\x, {200 - 60.1774*sqrt(1+pow((\x-500)/724.264,2))})
--(950,200)
--cycle;

\draw [thick, purple, fill, draw=purple] (50,100)
 to [out=80,in=280] (50,300)
 to [out=260,in=100] (50,100);
\draw[fill] (50,200) circle [radius=10];
\node[above right] at (50,320) {lens};
\draw [thick, red, fill, draw=red] (193,100) to [out=75,in=285] (193,300)--(193,100);
\draw[fill] (200,200) circle [radius=10];
\node[above] at (200,300) {cylinder};
\node[below] at (200,190) {$\mathbf{-a}$};
\draw [thick, red, fill, draw=red] (807,100) to [out=105,in=255] (807,300)--(807,100);
\draw[fill] (800,200) circle [radius=10];
\node[above] at (800,300) {cylinder};
\draw (0,200) --(1000,200);
\draw [<->] (210,200) --(790,200);
\draw[thick] (500,190) --(500,210);
\node[below] at (500,190) {$\mathbf{0}$};
\node[below] at (800,190) {$\mathbf{a}$};
\node[below] at (500,125) {\normalsize $d=f\sqrt{2}$};
\node[above] at (200,0) {\normalsize $w_i = \sqrt{2f/k}$};
\node[above] at (800,0) {\normalsize $w_o = \sqrt{2f/k}$};
\draw [->] (0,200) --(1000,200);
\node[above left] at (1000,200) {$z$};
\draw [red,->] (0,200) --(0,370);
\node[red, above] at (0,370) {$w_x$};
\draw [blue,->] (25,200) --(25,345);
\node[blue, above] at (35,340) {$w_y$};

\end{tikzpicture}

%% file: Figure3.tex
\begin{tikzpicture}[x=0.00485\linewidth,y=0.00485\linewidth]

\fill[blue, fill opacity=0.2, draw=blue, domain=40:160, variable=\x] 
(40,40)
--plot({\x}, {40 + 9.48683*sqrt(1+pow((\x-40+24)/72,2))})
--(160,40)
--(40,40)
--plot({\x}, {40 - 9.48683*sqrt(1+pow((\x-40+24)/72,2))})
--(160,40)
--cycle;

\fill[red, fill opacity=0.2, draw=red, domain=40:160, variable=\x] 
(40,40)
--plot({\x}, {40 + 5.14496*sqrt(1+pow((\x-40-35.2941)/21.1765,2))})
--(160,40)
--(40,40)
--plot({\x}, {40 - 5.14496*sqrt(1+pow((\x-40-35.2941)/21.1765,2))})
--(160,40)
--cycle;

\fill[red!50!blue, fill opacity=0.36, draw=purple, domain=10:40] 
(10,40)
--plot (\x, {40 + 8.3205*sqrt(1+pow((\x-40-36.9231)/55.3846,2))})
--(40,40)
--(10,40) 
--plot (\x, {40 - 8.3205*sqrt(1+pow((\x-40-36.9231)/55.3846,2))})
--(40,40)
--cycle;

\fill[red!50!blue, fill opacity=0.36, draw=purple, domain=160:190] 
(160,40)
--plot (\x, {40 + 6.7082*sqrt(1+pow((\x-160+108)/36,2))})
--(190,40)
--(160,40) 
--plot (\x, {40 - 6.7082*sqrt(1+pow((\x-160+108)/36,2))})
--(190,40)
--cycle;
\footnotesize
\draw [thick, purple, fill, draw=purple] (10,20) to [out=80,in=280] (10,60)
to [out=260,in=100] (10,20);
\draw[fill] (10,40) circle [radius=2];
\node[above] at (14,62) {lens};

\draw [thick, red, fill] (38,20) to [out=75,in=285] (38,60)--cycle;
\draw [thick, blue,fill] (38,18) 
to [out=75,in=285] (38,62)
--(42,62)
--(42,18)
--cycle;
\draw[fill] (40,40) circle [radius=2];
\node[above] at (40,62) {QP1};

\draw [thick, red, fill] (158,15) to [out=76,in=286] (158,65)--cycle;
\draw [thick, blue,fill] (158,13) 
to [out=76,in=286] (158,67)
--(162,67)
--(162,13)
--cycle;
\draw[fill] (160,40) circle [radius=2];
\node[above] at (160,67) {QP2};

\draw [<->] (43,40) --(157,40);
\node[above] at (100,40) {d};

\draw (0,40) --(200,40);
\footnotesize
\draw [->] (0,40) --(200,40);
\node[above left] at (200,40) {$z$};
\draw [red,->] (0,40) --(0,60);
\node[red, above] at (-1,60) {$w_x$};
\draw [blue,->] (4,40) --(4,70);
\node[blue, above] at (4,70) {$w_y$};

\draw [thick, red] (75.6,38) --(75.6,42);
\node[above, red] at (75.6,44) {$z_{0x}$};
\draw [thick, blue] (16,38) --(16,42);
\node[above, blue] at (18,42) {$z_{0y}$};

\end{tikzpicture}

%% file: Figure8.tex
\begin{tikzpicture}[x=0.0015\linewidth,y=0.0015\linewidth]
\footnotesize
\draw [thick, purple, fill, draw=purple] (50,-50) to [out=80,in=280] (50,50)
to [out=260,in=100] (50,-50);
\node[above, rotate = 0] at (70,65) {Hilbert beam};
\node[below, rotate = 0] at (50,-70) { $\diameter10~\mu$m};

\draw[fill, black] (215.5,0) circle [radius=2];

\draw [thick, purple, fill, draw=purple] (230,-50) to [out=80,in=280] (230,50)
to [out=260,in=100] (230,-50);
\node[above, rotate = 0] at (225,65) {demag};

\draw[thick, blue, fill] (252,-40) 
to (252, 40) 
to (262, 40) 
to (262,-40)--cycle;
\draw [thick, red, fill] (252,-30) to [out=75,in=285] (252,30)--cycle;
\node[above] at (257,35) {QP1};
\node[below] at (252,-70) { $\diameter0.7~\mu$m};

\draw[fill,lightgray] (257,0) circle [radius=2];

\draw[fill,red] (305,0) circle [radius=2];
\draw[fill,blue] (17,0) circle [radius=2];

\draw[thick, blue, fill] (372,-40) 
to (372, 40) 
to (382, 40) 
to (382,-40)--cycle;
\draw [thick, red, fill] (372,-30) to [out=75,in=285] (372,30)--cycle;
\node[above] at (372,35) {QP2};
\node[below] at (372,-70) { $\diameter1.5~\mu$m};

\draw[fill,lightgray] (377,0) circle [radius=2];

\draw [thick, purple, fill, draw=purple] (446,-50) to [out=80,in=280] (446,50)
to [out=260,in=100] (446,-50);
\node[above, rotate = 0] at (441,65) {mag};
\node[below] at (446,-45) { $\diameter2.4~\mu$m};

\draw[fill, black] (501.5,0) circle [radius=2];

\draw [thick, purple, fill, draw=purple] (562,-50) to [out=80,in=280] (562,50)
to [out=260,in=100] (562,-50);
\node[above, rotate = 0] at (562,65) {condenser};
\node[below] at (562,-70) { $\diameter2.6~\mu$m};

\draw [thick, purple, fill, draw=purple] (583.7,-50) to [out=80,in=280] (583.7,50)
to [out=260,in=100] (583.7,-50);
\node[above, rotate = 0] at (583.7,40) {objective};

\draw[fill] (586,0) circle [radius=2];
\node[right, rotate = 0] at (586,0) {focus};

\end{tikzpicture}